\documentclass[journal]{IEEEtran}

\usepackage{amsfonts,amssymb,color, graphicx, subcaption, amsmath}
\usepackage{caption}
\usepackage{cite}\usepackage{float}

\begin{document}
%

\title{Envelope Characterizations for Trajectories of the \'{C}uk Converter}


\author{\IEEEauthorblockN{Aleksandra Leki\'{c}\IEEEauthorrefmark{1},~\IEEEmembership{Senior Member,~IEEE}, and
Du\v{s}an Stipanovi\'{c}\IEEEauthorrefmark{2},~\IEEEmembership{Fellow,~IEEE}}
\thanks{Manuscript received ?? ?, ??; revised ?? ??, ??.
Corresponding author: A. Leki\'{c} (email: A.Lekic@tudelft.nl).

\IEEEauthorblockA{\IEEEauthorrefmark{1}Electrical Sustainable Energy, Faculty of Electrical Engineering, Mathematics and Computer Science, Delft University of Technology, Netherlands}
\IEEEauthorblockA{\IEEEauthorrefmark{2}Coordinated Science Laboratory, University of Illinois,
1308 West Main Street, Urbana, IL 61801, USA}}}

\markboth{Electrical Engineering}%
{Leki\'{c} \MakeLowercase{\textit{et al.}}: Perturbation Bound Estimation of the Sliding Mode Controlled \'{C}uk Converter}

\IEEEtitleabstractindextext{%
\begin{abstract}
In this paper, a complete characterization of the stability and the boundaries of the envelopes for trajectories of the \'{C}uk converter is provided in the cases of time and state-dependent switching. In particular, difference equations are developed and used to compute boundaries for the system's response envelopes as well as Lyapunov functions to show the system's stability in both continuous and discontinuous regimes of operation. Each case is illustrated by multiple simulations.
\end{abstract}

}

\maketitle

\IEEEdisplaynontitleabstractindextext

\IEEEpeerreviewmaketitle

\section{Introduction}
\IEEEPARstart
In this work, we analyze the stability and periodic behavior of the \'{C}uk converter according to time-dependent and state-dependent switching. \'{C}uk converter is one of the most sophisticated and complex DC-DC converters \cite{erickson2001fundamentals}. Its complexity can be illustrated by the fact that it has, in the continuous conduction modes, two marginally stable subsystems and four possible operating modes. Because of its complex transient behavior and the fact that it has continuous input and output currents, the \'{C}uk converter has been a frequent research topic of interest to many researchers over the years \cite{smedley1995dynamics,oppenheimer1996sliding, chen2012pi, martinez1998analysis}. Some of the works focus on controlling a single circuit variable to achieve proper control, as in \cite{smedley1995dynamics}, while more recent research accomplishments employ various control methods, such as sliding mode control \cite{oppenheimer1996sliding, chen2012pi,martinez1998analysis}, hysteresis control \cite{lekic2016hysteresis, 7752887}, or polytopic control \cite{8170285}.

By discretization of differential equations of the \'{C}uk converter's dynamic continuous-time models in different regimes \cite{kaagstrom1977bounds}, one obtains points on the boundaries of the envelopes for the systems' trajectories. This approach is suitable for analyzing the circuit's periodic behavior and for stability analysis using Lyapunov functions for discrete-time dynamic systems. Lyapunov function conditions are formulated as Linear Matrix Inequalities (LMIs) and can thus be computed using convex programming algorithms \cite{boy94}. This approach can be applied when the converter operates in continuous conduction modes (CCMs) \cite{lekic2016hysteresis} as well as in discontinuous modes \cite{maksimovic1991unified, 7752887}. While time-domain simulation and experimental results for the Ćuk converter under hysteresis control are reported in \cite{lekic2016hysteresis,7752887}, the present work provides a complete analytical characterization of the trajectory envelopes and stability conditions that govern those behaviors.
Unlike \cite{lekic2016hysteresis,7752887}, which focused on control design, the present work provides an envelope characterization valid across all four operating modes and both switching types, yielding closed-form DCM boundary conditions and a unified Lyapunov framework.

The paper itself is organized as follows. Section II provides an introduction to the circuit's operating modes and the state-space models. In Section III, the converter's behavior under time-dependent and state-dependent switching, is presented. In the first case, the converter's behavior is definitely periodic, and the stability conditions for all operating modes are determined. The same type of analysis is applied to state-dependent switching. Finally, the concluding remarks are provided in Section IV.

\section{Converter's operating modes and state-space model}
\begin{figure}[h!]
	\centering
	\includegraphics[width=\linewidth]{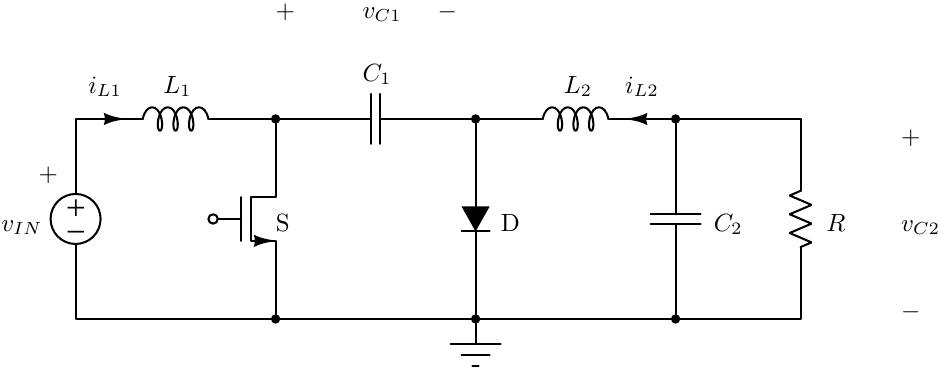}
	\caption{\'{C}uk converter.}
	\label{fig_cuk}
\end{figure}

The \'{C}uk converter, which is depicted in Fig. \ref{fig_cuk}, has two switches: switch S, realized as an unidirectional first quadrant switch (MOSFET or IGBT transistor); and switch D, realized as a diode.

This particular configuration of switches produces four operation modes of the converter, which can all be represented in the following linear-time invariant form \cite{lekic2016hysteresis, 7752887}:
\begin{equation}
\dot{\textbf{x}} = \textbf{A}_i \, \textbf{x} + \textbf{B}_i
\label{eq_dif}
\end{equation}
where $i \in \left\{1, 2, 3, 4\right\}$, since there are four operating modes. The state-vector is formed as $\textbf{x} = \begin{bmatrix}
i_{L1} & i_{L2} & v_{C1} & v_{C2}
\end{bmatrix}^T = \begin{bmatrix}
x_1 & x_2 & x_3 & x_4
\end{bmatrix}^T
$. The converter may operate in two CCMs. The first one is when switch S conducts, and the diode is turned off, which produces matrices $\textbf{A}_1$ and $\textbf{B}_1$ as follows:
\begin{equation}
\textbf{A}_1 = \begin{bmatrix}
0 & 0 & 0 & 0 \\
0 & 0 & \frac{1}{L_2} & \frac{1}{L_2} \\
0 & -\frac{1}{C_1} & 0 & 0  \\
0 & -\frac{1}{C_2} & 0 & -\frac{1}{RC_2}
\end{bmatrix}, \qquad
\textbf{B}_1 = \begin{bmatrix}
\frac{v_{IN}}{L_1} \\
0 \\
0 \\
0
\end{bmatrix}.
\end{equation}

The second CCM occurs when switch S is turned off and diode D conducts. In this mode, the state-space model is presented with state-space matrices $\textbf{A}_2$ and $\textbf{B}_2$ as follows:
\begin{equation}
\textbf{A}_2 = \begin{bmatrix}
0 & 0 & -\frac{1}{L_1} & 0 \\
0 & 0 & 0 & \frac{1}{L_2} \\
\frac{1}{C_1} & 0 & 0 & 0  \\
0 & -\frac{1}{C_2} & 0 & -\frac{1}{RC_2}
\end{bmatrix}, \quad
\textbf{B}_2 = \begin{bmatrix}
\frac{v_{IN}}{L_1} \\
0 \\
0 \\
0
\end{bmatrix}.
\end{equation}

Discontinuous capacitor voltage mode (DCVM) \cite{7752887, maksimovic1991unified} occurs when the voltage on capacitor $C_1$ reaches zero while switch S conducts. This forces diode D to turn on, which results in the following system's matrices $\textbf{A}_3$ and $\textbf{B}_3$:
\begin{equation}
\textbf{A}_3 = \begin{bmatrix}
0 & 0 & 0 & 0 \\
0 & 0 & 0 & \frac{1}{L_2} \\
0 & 0 & 0 & 0  \\
0 & -\frac{1}{C_2} & 0 & -\frac{1}{RC_2}
\end{bmatrix}, \qquad
\textbf{B}_3 = \begin{bmatrix}
\frac{v_{IN}}{L_1} \\
0 \\
0 \\
0
\end{bmatrix}.
\end{equation}

Discontinuous inductor current mode (DICM) \cite{7752887, maksimovic1991unified} occurs when the sum of inductor currents becomes zero, which forces diode D to turn off (note that switch S is already off). In this operating mode, both switches, S and D, are turned off, resulting in matrices $\textbf{A}_4$ and $\textbf{B}_4$ as follows:
\begin{equation}
\textbf{A}_4 = \begin{bmatrix}
0 & 0 & -\frac{1}{L_1 + L_2} & -\frac{1}{L_1 + L_2} \\
0 & 0 & \frac{1}{L_1 + L_2} & \frac{1}{L_1 + L_2} \\
\frac{1}{C_1} & 0 & 0 & 0  \\
0 & -\frac{1}{C_2} & 0 & -\frac{1}{RC_2}
\end{bmatrix}, \;
\textbf{B}_4 = \begin{bmatrix}
\frac{v_{IN}}{L_1 + L_2} \\
-\frac{v_{IN}}{L_1 + L_2} \\
0 \\
0
\end{bmatrix}.
\end{equation}

Clearly, matrices $\textbf{A}_1$, $\textbf{A}_3$ and $\textbf{A}_4$ are singular, meaning $\det \textbf{A}_i = 0$, while matrix $\textbf{A}_2$ is nonsingular. Given the format of ordinary differential equation (\ref{eq_dif}), the exact solution of the equation \cite{kaagstrom1977bounds, lekic2017stabilno}, under assumption that the initial condition is denoted as $\textbf{x}(0)$, is given by
\begin{equation}
\textbf{x}(t) = \mathrm{e}^{\textbf{A}_i \, t} \, \textbf{x}(0) + \int_0^t \mathrm{e}^{\textbf{A}_i \, \tau} \, \textbf{B} \, d\tau,
\label{eq_exact1}
\end{equation}
for $i \in \left\{1, 2, 3, 4\right\}$. If matrix $\textbf{A}_i$ is not singular, equation (\ref{eq_exact1}) can be simplified to
\begin{equation}
\textbf{x}(t) = \mathrm{e}^{\textbf{A}_i \, t} \, \textbf{x}(0) + \left(\mathrm{e}^{\textbf{A}_i \, t} - \textbf{I} \right) \, \textbf{A}_i^{-1} \, \textbf{B}.
\end{equation}

The converter, being a Qn-PWM quasi-resonant converter \cite{maksimovic1991unified}, can exhibit four types of switching during one period:
\begin{enumerate}
    \item If the converter is operating in CCMs operating subsystems: S = 1, D = 0 $\rightarrow$ S = 0, D = 1.
    \item If DICM occurs: S = 1, D = 0 $\rightarrow$ S = 0, D = 1 $\rightarrow$ S = 0, D = 0 (DICM).
    \item And if DCVM occurs: S = 1, D = 0 $\rightarrow$ S = 1, D = 1 (DCVM) $\rightarrow$ S = 0, D = 1.
    \item However, being a Qn-PWM type of converter, it has the possibility to operate as quasi resonant with state change: S = 1, D = 0 $\rightarrow$ S = 1, D = 1 (DCVM) $\rightarrow$ S = 0, D = 1 $\rightarrow$ S = 0, D = 0 (DICM).
\end{enumerate}

\section{Envelope switching characterization}
Under the assumption that the \'{C}uk converter operates in the CCMs, the difference equations can be derived as in equation \eqref{eq_diff}, where $\Delta_1$ and $\Delta_2$ are the time intervals for which the converter operates in the state where S is on and D off, and the state where S is off and D is on, respectively. In the case when the switching is time-dependent with period $T$ and duty-ratio $d$, the values for $\Delta_1$ and $\Delta_2$ are $\Delta_1 = dT$ and $\Delta_2 = (1 - d) \, T$. Vectors $\textbf{x}_1$ and $\textbf{x}_2$ denote the state vector values at the end of the corresponding operating modes, that is, $\textbf{x}_1$ after the first CCM and $\textbf{x}_2$ after the second. They are also representing the values on the envelope of the continuous-time state trajectories \cite{lekic2017stabilno}.

\begin{figure}[h!]
	\centering
	\includegraphics[width = \linewidth]{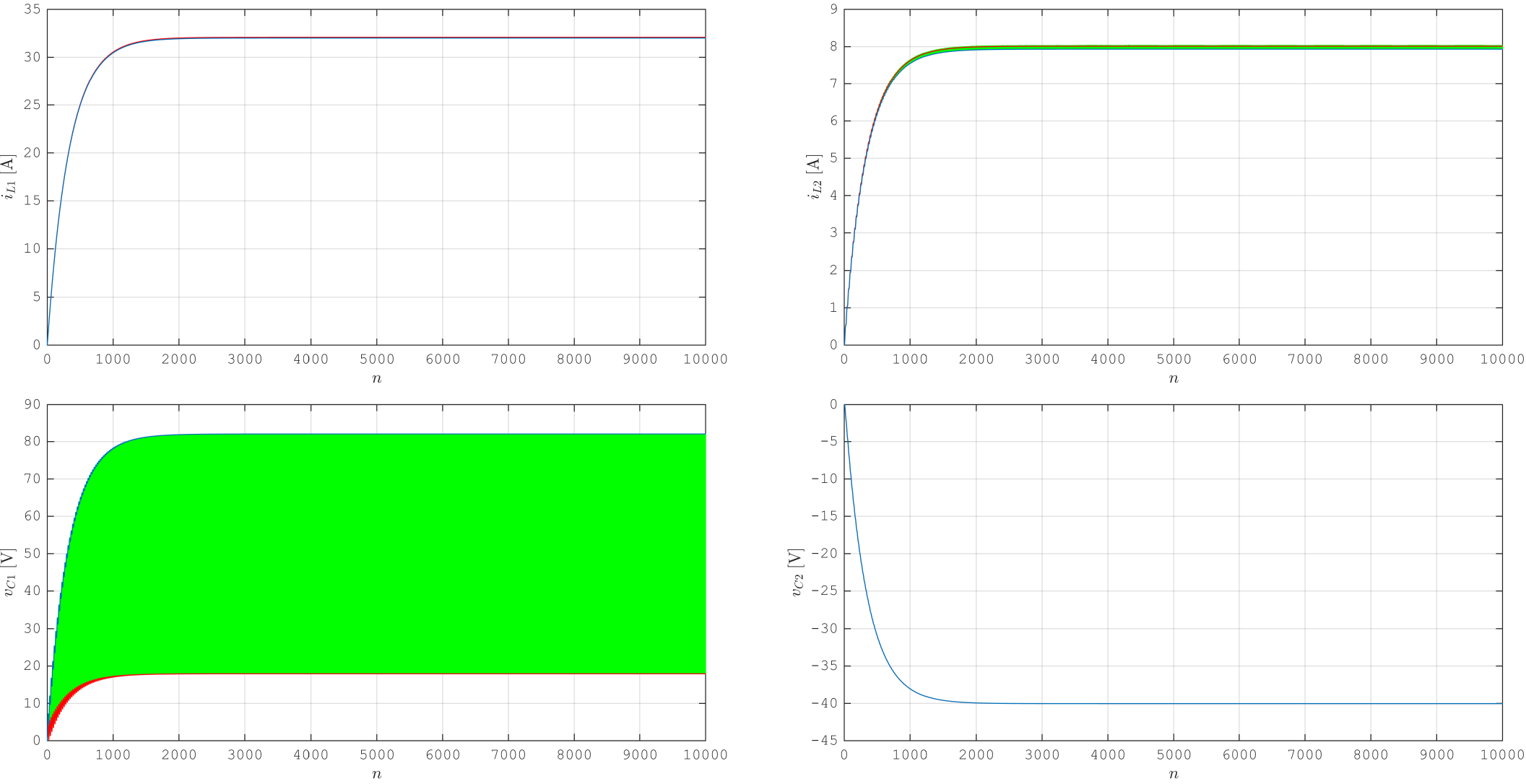}
	\caption{Transient of the \'{C}uk converter with $f_S = 100 \text{ kHz}$ and $d = 0.8$. Red line presents $\textbf{x}_1[n]$, blue line presents $\textbf{x}_2[n]$, and green line is a simulation from PLECS.}
	\label{fig_difference}
\end{figure}

In what follows the \'{C}uk converter's circuit elements are assumed to have particular values $L_1 = L_2 = 1 \text{ mH}$, $C_1 = 1 \; \mu \text{F}$, $C_2 = 20 \; \mu \text{F}$, $R = 5 \; \Omega$, and $v_{IN} = 10 \text{ V}$. In Figure \ref{fig_difference}, a transient simulation for the \'{Cuk} converter operating in CCM with switching frequency $f_S = 100 \text{ kHz}$ and duty-ratio $d = 0.8$ is presented. In Figure \ref{fig_difference}, the envelope border obtained through difference equations (\ref{eq_diff}) and diagrams obtained in PLECS for the same converter's parameters are plotted.

\begin{figure}[th!]
	\centering
	\includegraphics[width=3.5in]{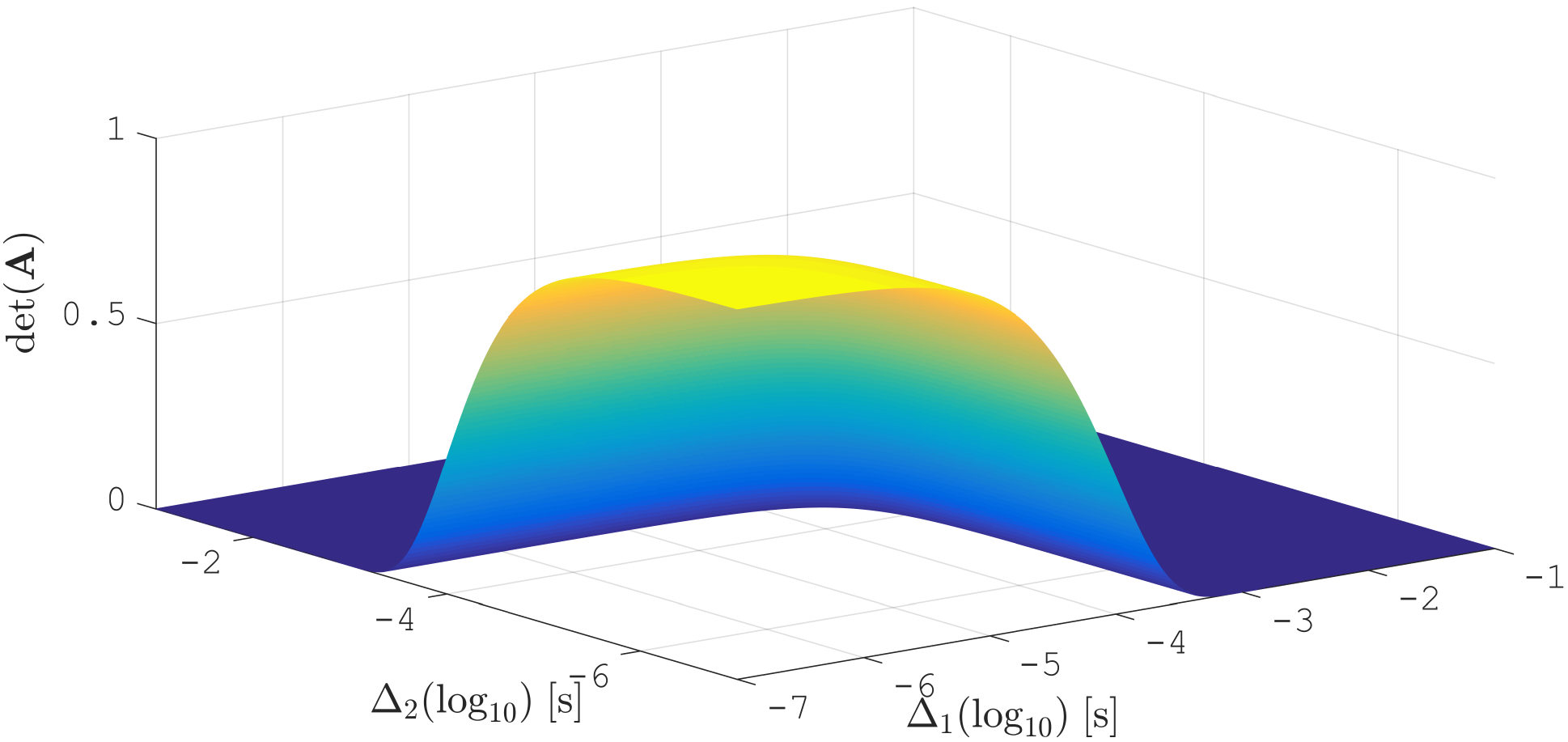}
	\caption{Dependance $\det \textbf{A}(\Delta_1, \Delta_2)$.}
	\label{fig_det}
\end{figure}
\begin{figure}[th!]
	\centering
	\includegraphics[width=3.5in]{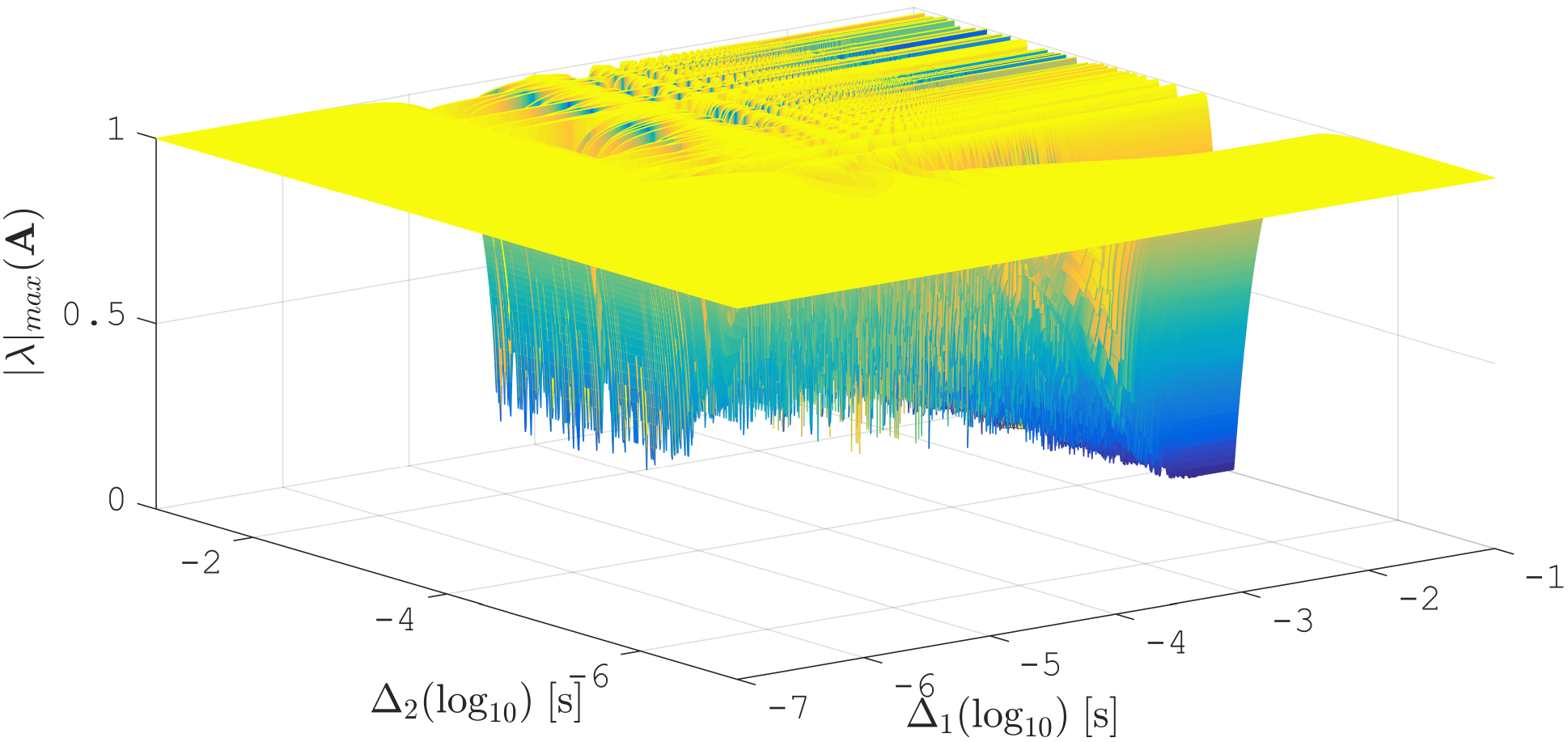}
	\caption{Dependance $|\lambda|_{max} \textbf{A}(\Delta_1, \Delta_2)$.}
	\label{fig_max}
\end{figure}
\begin{figure*}[ht!]
    \begin{equation}
\begin{array}{l}
\textbf{x}_1[k+1] = \mathrm{e}^{\textbf{A}_1 \, \Delta_1} \, \textbf{x}_2[k] + \int_0^{\Delta_1} \mathrm{e}^{\textbf{A}_1 \, \tau} \, \textbf{B}_1 \, d\tau 
= \mathrm{e}^{\textbf{A}_1 \, \Delta_1} \, \mathrm{e}^{\textbf{A}_2 \, \Delta_2} \, \textbf{x}_1[k] + \mathrm{e}^{\textbf{A}_1 \, \Delta_1} \, \left(\mathrm{e}^{\textbf{A}_2 \, \Delta_2} - \textbf{I}\right) \; \textbf{A}_2^{-1} \, \textbf{B}_2 + \int_0^{\Delta_1} \mathrm{e}^{\textbf{A}_1 \, \tau} \, \textbf{B}_1 \, d\tau, \\
\textbf{x}_2[k+1] = \mathrm{e}^{\textbf{A}_2 \, \Delta_2} \, \textbf{x}_1[k+1] + \left(\mathrm{e}^{\textbf{A}_2 \, \Delta_2} - \textbf{I}\right) \; \textbf{A}_2^{-1} \, \textbf{B}_2
= \mathrm{e}^{\textbf{A}_2 \, \Delta_2} \, \mathrm{e}^{\textbf{A}_1 \, \Delta_1} \, \textbf{x}_2[k] + \mathrm{e}^{\textbf{A}_2 \, \Delta_2} \, \int_0^{\Delta_1} \mathrm{e}^{\textbf{A}_1 \, \tau} \, \textbf{B}_1 \, d\tau + \left(\mathrm{e}^{\textbf{A}_2 \, \Delta_2} - \textbf{I}\right) \; \textbf{A}_2^{-1} \, \textbf{B}_2,
\end{array}
\label{eq_diff}
\end{equation}   
\begin{equation}
\begin{array}{ll} \vspace*{2mm}
\textbf{x}_1[k+1]  = &\mathrm{e}^{\textbf{A}_1 \, \Delta_1} \, \textbf{x}_3[k] + \int_0^{\Delta_1} \mathrm{e}^{\textbf{A}_1 \, \tau} \, \textbf{B}_1 \, d\tau, \\
\textbf{x}_2[k+1] = &\mathrm{e}^{\textbf{A}_3 \, \Delta_3} \, \mathrm{e}^{\textbf{A}_1 \, \Delta_1} \, \textbf{x}_3[k] + \mathrm{e}^{\textbf{A}_3 \, \Delta_3} \, \int_0^{\Delta_1} \mathrm{e}^{\textbf{A}_1 \, \tau} \, \textbf{B}_1 \, d\tau  + \int_0^{\Delta_3} \mathrm{e}^{\textbf{A}_3 \, \tau} \, \textbf{B}_3 \, d\tau, \\
\textbf{x}_3[k+1] = &\mathrm{e}^{\textbf{A}_2 \, \Delta_2} \, \mathrm{e}^{\textbf{A}_3 \, \Delta_3} \, \mathrm{e}^{\textbf{A}_1 \, \Delta_1} \, \textbf{x}_3[k] + \mathrm{e}^{\textbf{A}_2 \, \Delta_2} \, \mathrm{e}^{\textbf{A}_3 \, \Delta_3} \, \int_0^{\Delta_1} \mathrm{e}^{\textbf{A}_1 \, \tau} \, \textbf{B}_1 \, d\tau + \mathrm{e}^{\textbf{A}_2 \, \Delta_2} \, \int_0^{\Delta_3} \mathrm{e}^{\textbf{A}_3 \, \tau} \, \textbf{B}_3 \, d\tau + \left(\mathrm{e}^{\textbf{A}_2 \, \Delta_2} - \textbf{I}\right) \; \textbf{A}_2^{-1} \, \textbf{B}_2,
\end{array} 
\label{eq_diff_dcvm}
\end{equation}
\hrule

	\centering
    \vspace*{3mm}
	\includegraphics[width=0.75\textwidth]{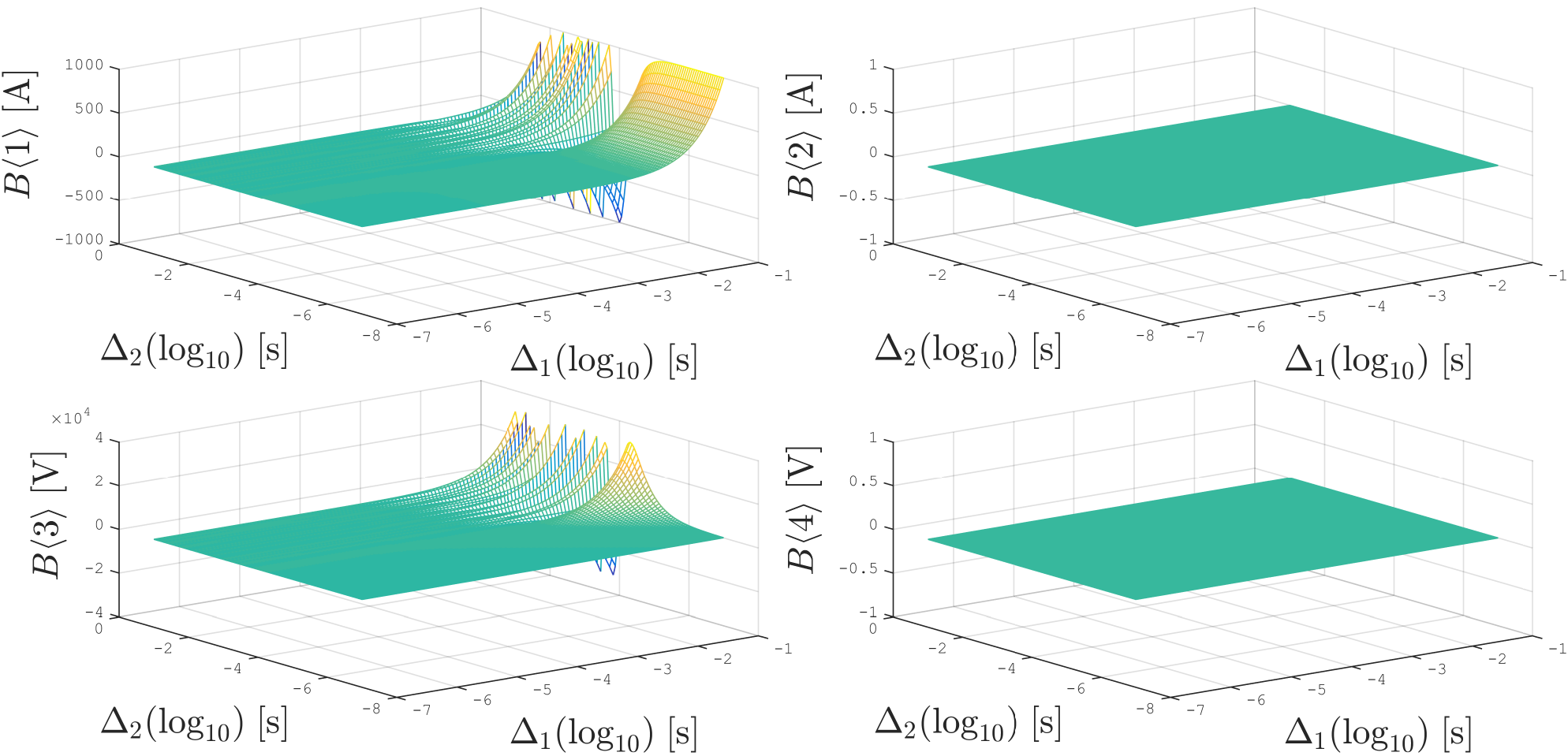}
	\caption{Dependance $ \textbf{B}(\Delta_1, \Delta_2)$.}
	\label{fig_b}
\end{figure*}

Figs. \ref{fig_det} and \ref{fig_max} provide diagrams showing the dependence of the system matrix's determinant and its maximal absolute eigenvalue on $(\Delta_1, \Delta_2)$. Clearly, these diagrams show that the system is stable for every value of $\Delta_1$ and $\Delta_2$.


In Fig. \ref{fig_b}, diagrams for the matrix function $\textbf{B}(\Delta_1, \Delta_2)$, are depicted. As can be seen, variables $x_1$ and $x_3$ are the critical ones because their values, when $\Delta_1$ is large are also large, thereby forcing the system into discontinuous modes.

For the converter operating in DCVM, difference equations are given in \eqref{eq_diff_dcvm}, where $\textbf{x}_1$ represent the value of the state variables after first CCM is over (that is, when S = 1 and D = 0), $\textbf{x}_2$ takes final values after DCVM mode, and $\textbf{x}_3$ after the second CCM (again, when S = 0 and D = 1). The simulations provide maximum absolute eigenvalue $\max(|\lambda|_{max}(\textbf{A}(\Delta_1, \Delta_2, \Delta_3))) < 1$ and $\max(\det(\textbf{A}(\Delta_1, \Delta_2, \Delta_3))) = 0.9970$ which imply that the system is at least boundary stable. The same analysis can be performed for DICM and DQRM regimes.

The \'{C}uk converter described with difference equations (\ref{eq_diff}) has the same Lyapunov function to establish stability for both $\textbf{x}_1$ and $\textbf{x}_2$. In our analysis, we will focus on the difference equation for $\textbf{x}_2$ and we will denote $\textbf{y} \equiv \textbf{x}_2 - \overline{\textbf{x}}$, where $\overline{\textbf{x}}$ denotes the vector of equilibrium values (steady-state values if there is a steady-state). The Lyapunov function difference equation for this system is given by
\begin{equation}
V(\textbf{y}[k+1]) = \textbf{y}^T[k+1] \, \textbf{P} \, \textbf{y}[k+1]
\end{equation}
which results in
\begin{equation}
\begin{array}{ll}
\Delta V(\textbf{y}[k]) = & \textbf{y}^T[k] \; \left(A^T[k] \, \textbf{P} \, \textbf{A}[k] - \textbf{P}\right) \; \textbf{y}[k] \\
&  + 2 \, \textbf{B}^T[k] \, \textbf{P} \, \textbf{A}[k] \, \textbf{y}[k] + \textbf{B}^T[k] \, \textbf{P} \, \textbf{B}[k].
\end{array}
\end{equation}
This equation can be simplified by introducing the following notation:
\begin{equation}
\textbf{Q}[k] = A^T[k] \, \textbf{P} \, \textbf{A}[k] - \textbf{P}.
\end{equation}

In the discrete case, the LMIs transform to
\begin{equation}
\begin{array}{ll}
\textbf{A}^T \, \textbf{P} \, \textbf{A} - \textbf{P} \prec 0, \\
\textbf{P} \succ 0,
\end{array}
\label{eq_lmi_d}
\end{equation}
where $\textbf{A} = \mathrm{e}^{\textbf{A}_2 \Delta_2} \, \mathrm{e}^{\textbf{A}_1 \Delta_1}$. This matrix $\textbf{P}$ also works for the discrete-time difference equations by establishing stability wherever the system is operating in CCMs by satisfying inequalities (\ref{eq_lmi_d}).

For example, for the simulated case shown in Fig. \ref{fig_difference} and corresponding matrix $\textbf{P}$
\begin{equation}
\textbf{P} = 10^{-8} \cdot
\begin{bmatrix}
    0.1209  &  0  &  0  &  0 \\
    0  &  0.1210  &  0  &  0.0003 \\
    0  &  0  &  0.0001  &  0 \\
    0  &  0.0003  &  0  &  0.0025
\end{bmatrix}
\label{eq_calc_p}
\end{equation}
the obtained matrix $\textbf{Q}$ is positive definite 
\begin{equation}
\textbf{Q} =  10^{-11} \cdot \begin{bmatrix}
    0.0003  &  0.0030  &  0.0000  &  0.0003 \\
    0.0030  &  0.2879  &  0.0013  & -0.0172 \\
    0.0000  &  0.0013  &  0.0000  & -0.0024 \\
    0.0003  & -0.0172  & -0.0024  &  0.4536
\end{bmatrix}. \label{eq_calc_q}
\end{equation}
The matrix $\textbf{P}$ in \eqref{eq_calc_p} is computed for the operating point shown in Fig. 2, confirming CCM operation through positive definiteness of $\textbf{Q}$ in \eqref{eq_calc_q}.

In what follows, the stability and periodical behavior of the \'{C}uk converter operating in CCMs will be analyzed. Analyses are made according to time-dependent and state-dependent switching. It is shown that the Lyapunov function for a discrete-time system, evaluated at three points, can be used to determine the operating mode. In the CCM, all eigenvalues of $\texttt{Q}$ are inside the circle of radius 1. However, with a determined Lyapunov function in the DCM matrix $\textbf{Q}$, it has at least one eigenvalue outside the unit circle.  

\subsection{Time-dependent switching}
\begin{figure}[h!]
	\centering
	\includegraphics[width = \linewidth]{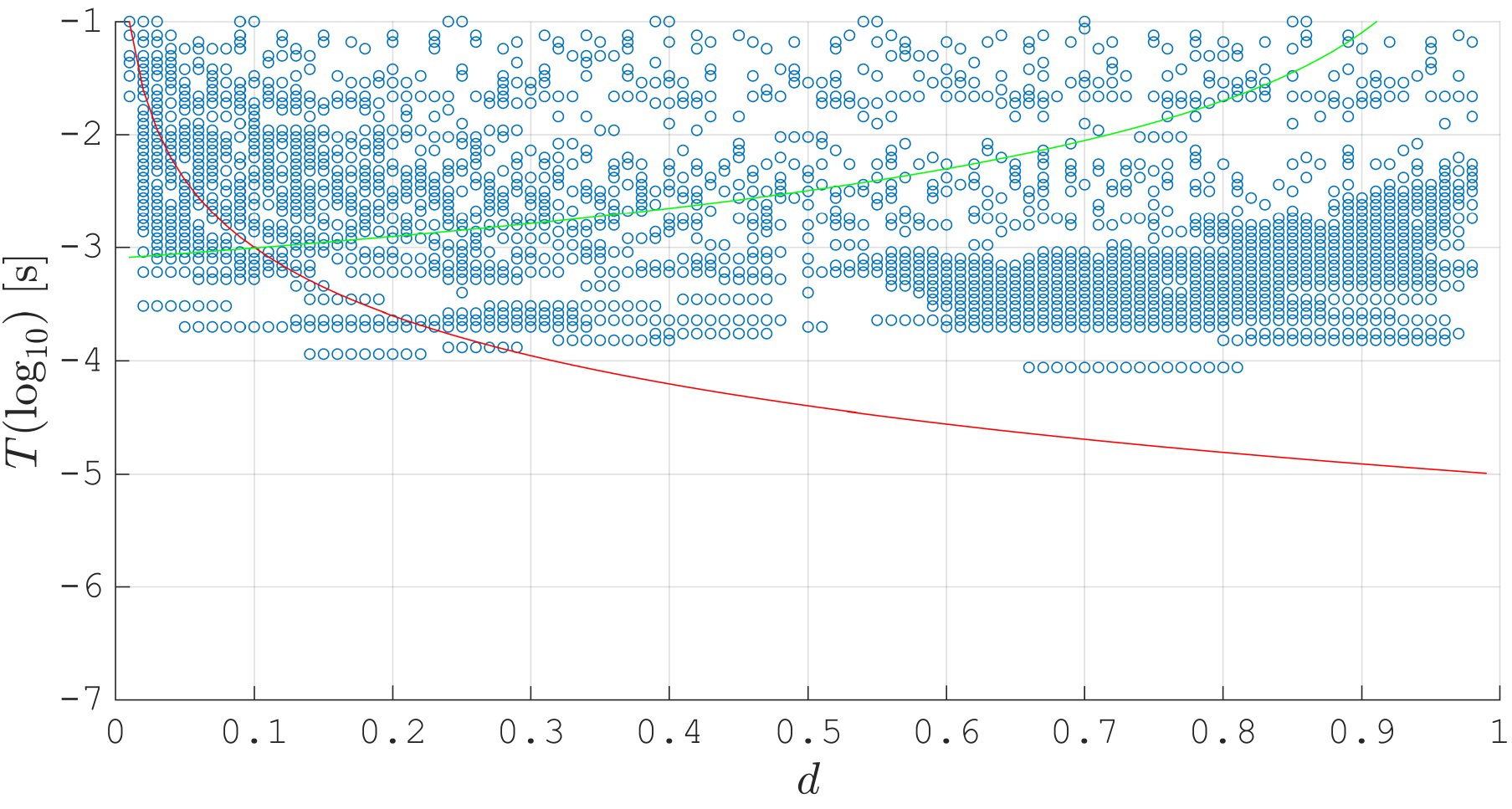}
	\caption{Diagram representing DCM border according to $\lambda(\textbf{Q}(d, T)) < 0$.}
	\label{fig_stab}
    \vspace*{-5mm}
\end{figure}

Periodic behavior of the converter for the time-dependent switching is induced by controlling it with a constant switching period $T$. Lyapunov function for some period $T$ is determined by setting duty-ratio at the beginning of the interval to be $d_1 = 0.01$, at the end of the interval $d_2 = 0.99$, and in the middle $d_3 = 0.5$, we obtain the following LMIs:
\begin{equation}
\begin{array}{ll}
\textbf{A}_{x1}^T \, \textbf{P} \, \textbf{A}_{x1} - \textbf{P} \prec 0, \\
\textbf{A}_{x2}^T \, \textbf{P} \, \textbf{A}_{x2} - \textbf{P} \prec 0, \\
\textbf{A}_{x3}^T \, \textbf{P} \, \textbf{A}_{x3} - \textbf{P} \prec 0, \\
\textbf{P} \succ 0,
\end{array} \label{eq:discrete_lims}
\end{equation}
where $\textbf{A}_{xi} = \mathrm{e}^{\textbf{A}_2 \, (1-d_i) \, T} \, \mathrm{e}^{\textbf{A}_1 d_i T}$. Now, we can vary duty-ratio with fixed period $T$ and check if the same Lyapunov matrix  works for $\textbf{A}_x = \mathrm{e}^{\textbf{A}_2 \, (1-d) \, T} \, \mathrm{e}^{\textbf{A}_1 d T}$. In Fig. \ref{fig_stab}, values for $(d, T)$ for which matrix
\begin{equation}
\textbf{Q} = -\textbf{A}_x^T \, \textbf{P} \, \textbf{A}_x + \textbf{P}, \label{eq:q}
\end{equation}
has at least one negative eigenvalue are depicted with dots. Focusing on paper \cite{maksimovic1991unified}, for switching period
\begin{equation}
T \ge \dfrac{2 R C_1}{d^2} = \dfrac{a}{d^2}, \quad a = 2RC_1
\label{eq_dcvm}
\end{equation}
the converter operates in DCVM and for
\begin{equation}
T \ge \dfrac{2 (L_1 + L_2)}{R (1 - d)^2} = \dfrac{b}{(1-d)^2}, \quad b = \dfrac{2 (L_1 + L_2)}{R}
\label{eq_dicm}
\end{equation}
in DICM. In Fig. \ref{fig_stab}, the border for operation in DCVM (\ref{eq_dcvm}) is depicted in red, and similarly, in green for DICM (\ref{eq_dicm}). Dots represent values for $(d, T)$ for which the system has at least one negative eigenvalue (that is, $\lambda(\textbf{Q}(d, T)) < 0$).

\begin{figure}[h!]
	\centering
	\includegraphics[width = 3.5in]{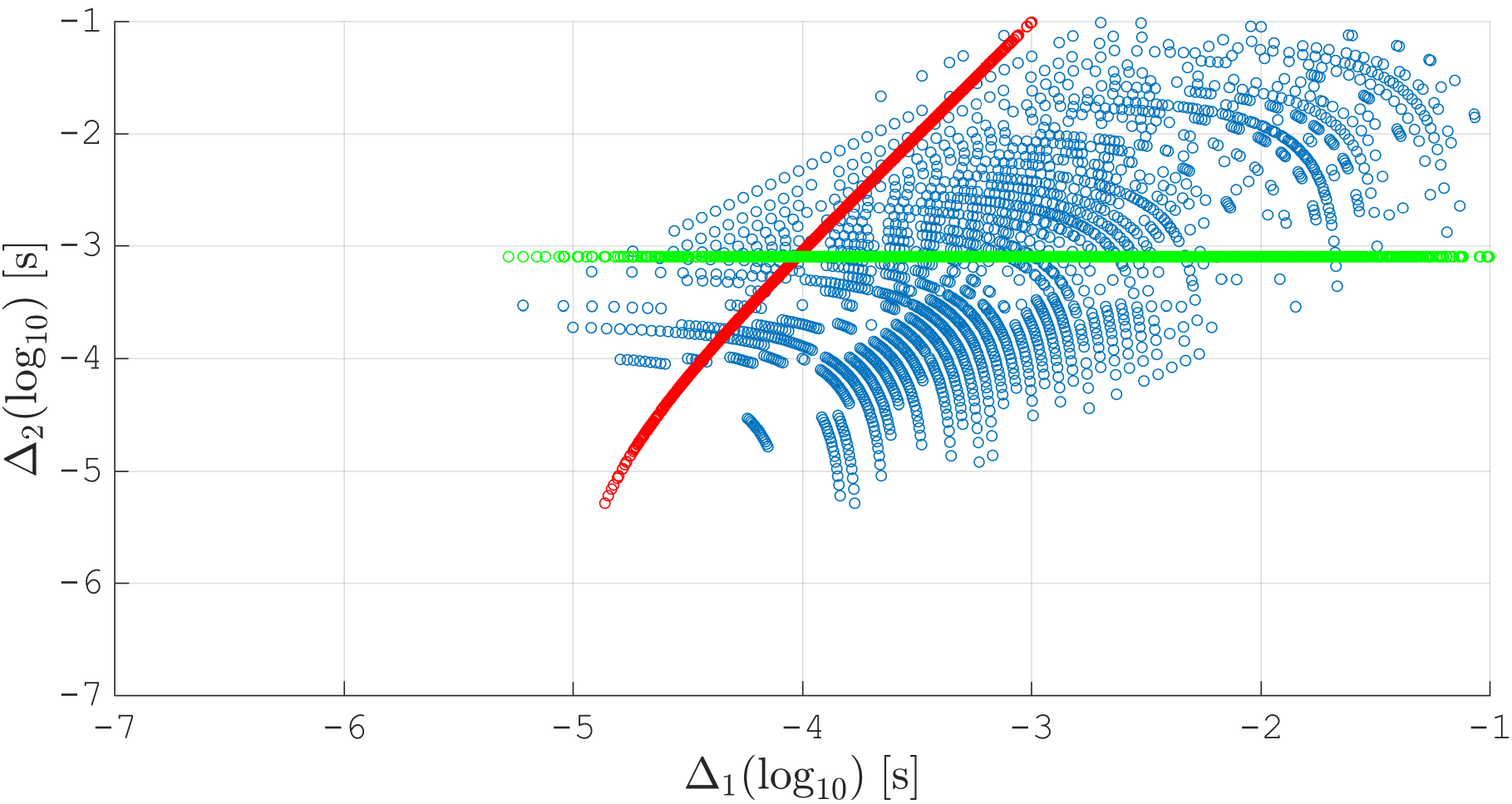}
	\caption{Diagram representing DCM border according to $\lambda(\textbf{Q}(\Delta_1, \Delta_2)) < 0$.}
	\label{fig_stab2}
\end{figure}

It can be observed that the area in which the converter operates in CCMs according to matrix $\textbf{Q}$ isn't covered with blue dots (Fig. \ref{fig_stab}), except in several cases near the border, where instability of matrix $\textbf{Q}$ is caused by numerical errors. In Fig. \ref{fig_stab2}, the same analysis is provided with respect to $(\Delta_1, \Delta_2)$. The condition for entering DCVM is
\begin{equation}
\Delta_1 \ge \dfrac{a + \sqrt{a^2 + 4 \, a \, \Delta_2}}{2},
\label{eq_d_dcvm}
\end{equation}
which is depicted in Fig. \ref{fig_stab2} as the red border line, and for DICM
\begin{equation}
\Delta_2 \ge \dfrac{b + \sqrt{b^2 + 4 \, b \, \Delta_1}}{2},
\label{eq_d_dicm}
\end{equation}
as the green border line. The same effect can be seen in Fig. \ref{fig_stab2}.

Given a desired switching period $T$ and duty ratio $d$, the designer can verify CCM operation by checking whether the operating point falls below both boundary curves. If the point approaches the red border, DCVM onset is imminent and either $T$ should be reduced or d adjusted. The green border similarly warns of DICM entry.

\vspace*{-5mm}
\subsection{State-dependent switching}
State-dependent switching corresponds to hysteresis control, where the switching instant is determined by the state crossing a threshold rather than a fixed clock. The envelope bounds derived here provide guaranteed ripple limits under this control strategy without requiring time-domain simulation.

\begin{figure}[h!]
	\centering
	\includegraphics[width = 3.5in]{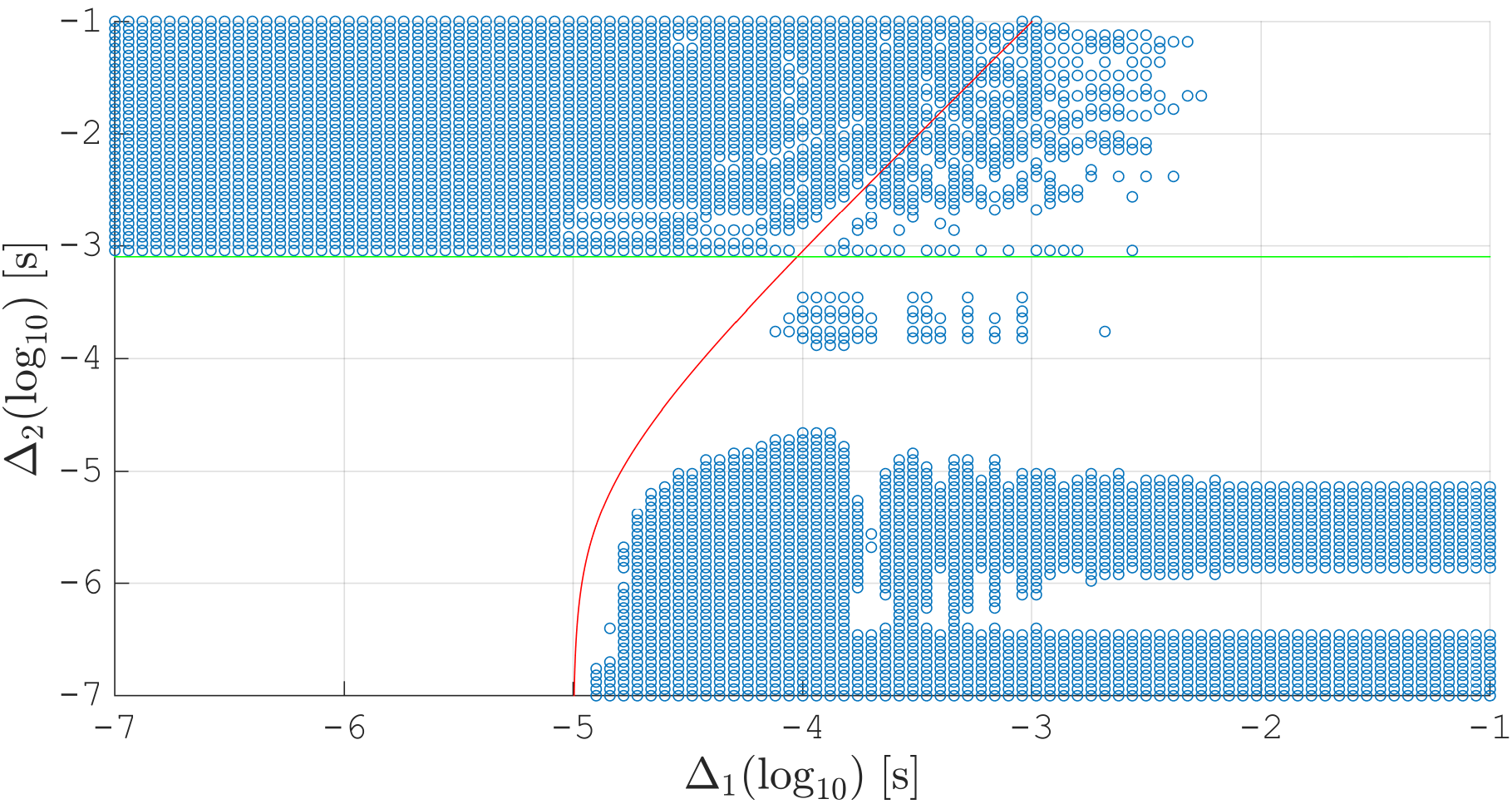}
	\caption{Diagram representing DCM border according to $\lambda(\textbf{Q}(\Delta_1, \Delta_2)) < 0$.}
	\label{fig_stab3}
\end{figure}

State-dependent switching provides a global result because it can give a nonhomogeneous interval duration. However, it yields results similar to those of time-dependent switching. Stability of the \'{C}uk converter's behavior is determined by $\left| \lambda_{max} \left(\mathrm{e}^{\textbf{A}_2 \Delta_2} \, \mathrm{e}^{\textbf{A}_1 \Delta_1} \right) \right| < 1$ which can be observed from Fig. \ref{fig_max}. However, a formal stability condition can be formulated and tested using the LMIs from equation \eqref{eq:discrete_lims}, where $\textbf{A}_{xi} = \mathrm{e}^{\textbf{A}_2 \, \Delta_2} \, \mathrm{e}^{\textbf{A}_1 \Delta_{1i}}$, $\Delta_2$ has some fixed value, and thus matrix $\textbf{P}$ is computed for three values of $\Delta_1$ being
\begin{equation}
\begin{array}{l}
\Delta_{11} = \frac{a + \sqrt{a^2 + 4 \, a \, \Delta_2}}{2}, \\
\Delta_{12} = \left\{
\begin{array}{ll}
\frac{\Delta^2_2}{b} - \Delta_2, & \quad \frac{\Delta^2_2}{b} - \Delta_2 > 0 \\
\min(\Delta_1), & \quad \frac{\Delta^2_2}{b} - \Delta_2 \le 0,
\end{array} \right. \\
\Delta_{13} = \sqrt{\Delta_{11} \, \Delta_{12}}
\end{array}.
\end{equation}
Pairs $(\Delta_1, \Delta_2)$ for which matrix
\begin{equation}
\textbf{Q} = -\textbf{A}_x^T \, \textbf{P} \, \textbf{A}_x + \textbf{P} \label{eq:q2}
\end{equation}
has at least one eigenvalue of $\textbf{A}_{x} = \mathrm{e}^{\textbf{A}_2 \, \Delta_2} \, \mathrm{e}^{\textbf{A}_1 \Delta_{1}}$ is negative, are depicted with blue circles in Fig. \ref{fig_stab3}. It can be seen that Fig. \ref{fig_stab3} shows the same property as Fig. \ref{fig_stab2}.

Fig. \ref{fig_stab3} shows that the Lyapunov function provides information about converter operating conditions. Clearly, when the converter operates in CCM, the Lyapunov function indicates that the system is stable. In the case when \'{C}uk operates in some DCM, the Lyapunov function calculated for the discrete-time system operating in CCMs provides information that the system is not stable. This can be used to check if the system operates in CCMs. However, the converter in DCM is stable, as shown in (\ref{eq_diff_dcvm}).

To summarize, the matrix $\textbf{Q}$ computed in equations \eqref{eq:q} and \eqref{eq:q2} can serve as an online indicator of the operating mode. If during operation the eigenvalues of $\textbf{Q}$ become negative, this signals transition toward DCM before it occurs, providing a basis for a supervisory controller to adjust $d$ or T preemptively. This requires only matrix-vector multiplications at each switching instant and is computationally feasible in real-time embedded implementations.

\section{Conclusion}
In this paper, we analyzed the stability of the \'{C}uk converter according to time-dependent and state-dependent switching. Ordinary differential equations describing each converter subsystem are discretized, yielding a set of discrete equations that provide an exact transient envelope. For the discrete circuit model, stability and periodic behavior are provided for two switching types: time- and state-dependent. For each switch, we constructed a discrete Lyapunov function and determined conditions for operating in discontinuous conduction mode.

The derived envelope bounds directly yield worst-case estimates of inductor current ripple and output voltage ripple as functions of switching period and duty ratio, enabling analytical design without time-domain simulation. The Lyapunov-based DCM detection condition provides a computationally lightweight criterion suitable for real-time supervisory control.

\bibliographystyle{IEEEtran}      
\bibliography{references}   

\end{document}